\documentclass{article}

\usepackage{arxiv}

\usepackage[utf8]{inputenc} % allow utf-8 input
\usepackage[T1]{fontenc}    % use 8-bit T1 fonts
\usepackage{hyperref}       % hyperlinks
\usepackage{url}            % simple URL typesetting
\usepackage{booktabs}       % professional-quality tables
\usepackage{amsfonts}       % blackboard math symbols
\usepackage{nicefrac}       % compact symbols for 1/2, etc.
\usepackage{microtype}      % microtypography
\usepackage{lipsum}
\usepackage{amsmath}
\usepackage{graphicx}
\usepackage{amssymb}
\usepackage{float}

\begin{document}

\begin{flushleft}
{\Large
\textbf\newline{\textbf{Transfer function asymmetry in Fabry-P\'erot based optical pressure sensors}} %Fabry-P\'erot based photoacoustic system with Adaptive Optics correction for optimized coupling into the cavity}}
}
\newline
% authors go here:
\\
Jakub Czuchnowski\textsuperscript{1,\S},
Robert Prevedel\textsuperscript{1,*}
\\
\bigskip
\textsuperscript{1} Cell Biology and Biophysics Unit, European Molecular Biology Laboratory, Heidelberg, Germany
\\
\textsuperscript{\S} Collaboration for joint PhD degree between EMBL and Heidelberg University, Faculty of Biosciences, Germany
\\
\bigskip
* robert.prevedel@embl.de

\end{flushleft}

\begin{abstract}
Optical resonators are one of the most promising optical devices for manufacturing high-performance pressure sensors for photoacoustic imaging. Among these, Fabry-P\'erot (FP) based pressure sensors have been successfully used for a multitude of applications. However, critical performance aspects of FP based pressure sensors have not been extensively studied, including the effects system parameters such as beam diameter and cavity misalignment have on transfer function shape. Here, we discuss the possible origins of the transfer function asymmetry, the impact it has on measurement sensitivity as well as ways to correctly estimate the FP pressure sensitivity under practical experimental conditions.

\end{abstract}

\section{Introduction}
Optical resonators have emerged as powerful optical devices for next generation, high-performance pressure sensors for photoacoustic imaging \cite{Wissmeyer:18}. Among the possible implementations, Fabry-P\'erot (FP) based pressure sensors have proven especially promising and have been successfully used for bio-imaging applications \cite{Zhang:08,Jathoul:15}. However, a comprehensive picture on the performance of FP pressure sensors is still missing. Recent work in the field focused on the effects that surface roughness \cite{marques2021studying}, mirror parallelism \cite{marques2021analysing} as well as optical aberrations \cite{Czuchnowski:20} have on the shape and sensitivity of the inteferometer transfer function (ITF) of a FP pressure sensor. Here we extend this work by investigating the often present asymmetry in the ITF of FP pressure sensors as well as explaining the origins of this phenomenon and methods for tackling it's effects under experimental conditions.

This property, characteristic for FPs illuminated with focused beams, has not been extensively described or studied in the literature, yet has important implications for accurately quantifying the sensitivity of such sensors which would allow for more standardised comparisons between FP sensors with different cavity parameters and application regimes. Also, this becomes especially crucial for recently described approaches for improving FP sensitivity by Adaptive Optics (AO) enhancement of the interrogation beam properties \cite{Czuchnowski:20} where the experimentally quantified sensitivity is used as a feedback metric for the optimisation. Here, we discusses the origin of ITF asymmetry in FP pressure sensors, the impact of common FP imperfections on this asymmetry as well as propose a model for fitting asymmetric ITFs that improves the accuracy of the sensitivity measurement.

\begin{figure*}[h]
\includegraphics[width=16cm]{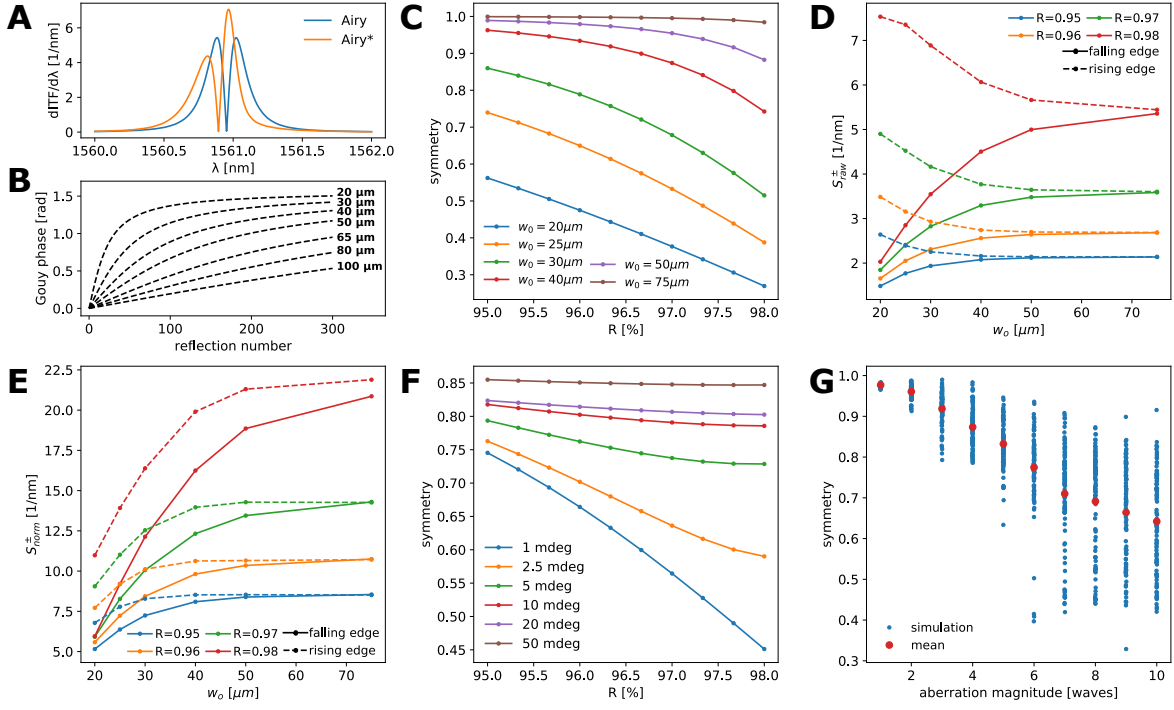}
\centering
\caption{\textbf{A} Comparison of the ITF derivative between a normal Airy ITF and an Airy ITF modified with a Gouy phase term (Airy*) \textbf{B} Gouy phase accumulation across different reflections. \textbf{C} Changes of the ITF asymmetry with beam radius ($w_0$) and mirror reflectivity ($R$). \textbf{D} Dependence of the raw optical sensitivity ($S_{raw}^\pm$) on beam radius. \textbf{E} Relation of the normalised optical sensitivity ($S_{norm}^\pm$) to beam diamter. \textbf{F} ITF asymmetry dependence on FPI wedge angle and mirror reflectivity ($R$) for a $w_0=25\ \mu m$ beam. \textbf{G} Effects of optical aberration magnitude on the ITF asymmetry. For all panels: cavity thickness - $20 \ \mu m$, refractive index inside the cavity - 1.639, simulated wavelength range 1560-1562 nm.} 
\label{fig:1}
\end{figure*}

%-----------------------------------
%	SECTION 1
%-----------------------------------

\section{Origins of transfer function asymmetry}

The interferometer transfer function (ITF) for a FP pressure sensor can be defined as:

\begin{equation}
ITF (\lambda) = \iint \limits_{A} E_{FPI}(x,y,\lambda)^*E_{FPI}(x,y,\lambda)dA 
\label{eq:I_FPI}
\end{equation}

where, in general:

\begin{equation}
E_{FPI}(x,y,\lambda) ={r_1^L}E(x,y,0,\lambda) + \sum \limits_{k=1}^{\infty}\beta_k E(x,y,z_k,\lambda')
\label{eq:Efpi}
\end{equation}

describes the electric field propagating inside the cavity with $\beta_k={(t_1)}^2{(r_1^R)}^{k-1}{(r_2^L)}^{k}$, $z_k=2l_0k$, $l_0$ the cavity length, $\lambda'$ the effective wavelength inside the cavity medium, $t_1$ the amplitude transmission coefficient for the first mirror, $r_1^{R/L}$ the amplitude reflection coefficients for the first mirror on the right respectively, and $r_2^L$ the amplitude reflection coefficient for second mirror for the left side. The particular differences in the ITF arise from either different coefficients or a different form of the propagating electric field ($E(x,y,z_k,\lambda)$). In the most basic case of an Airy ITF we have:

\begin{equation}
E_{Airy}(x,y,z,\lambda)=E_0\exp(-2i\pi z/\lambda)
\end{equation}

This gives rise to the canonical ITF shape that is strongly symmetric in wavelength space (and fully symmetric in wave-number space, \textbf{Figure \ref{fig:1}A}) \cite{Varu:14}. However, realistic FP pressure sensors illuminated with Gaussian beams show an ITF asymmetry which can be defined as:

\begin{equation}
    symmetry=\bigg|\frac{\min(dITF(\lambda)/d\lambda)}{\max(dITF(\lambda)/d\lambda)}\bigg|
\end{equation}

Where an symmetry value of 1 (0) defines a perfectly symmetric (asymmetric) ITF. This asymmetry can arise from multiple sources, but on the most fundamental level it is a consequence of Gouy phase accumulation \cite{Varu:14}, which for an arbitrary Laguerre-Gauss beam ($LG_{lp}$) can be described as:

\begin{equation}
    \phi_{Gouy}=(2p+|l|+1)arctan(z/z_r)
    \label{eq:Gouy}
\end{equation}

where $z_r=\pi w_0^2/\lambda$, for a Gaussian beam ($LG_{00}$) this reduces to $\phi_{Gouy}=arctan(z/z_r)$. To show this we use a modified Airy ITF where we add a Gouy term to the plane wave phase to reproduce an asymmetrical ITF (\textbf{Figure \ref{fig:1}A}):

\begin{equation}
E_{Airy*}(x,y,z,\lambda)=E_0\exp(-2i\pi z/\lambda+i\phi_{Gouy})
\end{equation}

As the Gouy phase accumulation is dependent on the beam divergence/diameter (\textbf{Figure \ref{fig:1}B}), a larger  divergence leads to an increased asymmetry which can be seen in \textbf{Figure \ref{fig:1}C}. Additionally, as the effects of divergence can be partially overshadowed by the limited finesse of the cavity, the asymmetry is also dependent on the mirror reflectivity with stronger reflecting mirrors displaying enhanced asymmetry. This asymmetry has interesting properties when it comes to the optical sensitivity of the FP pressure sensor. Counter-intuitively, the raw optical sensitivity ($S_o^{raw}$) increases on the rising edge with decreasing beam radius ($w_0$, \textbf{Figure \ref{fig:1}D}):

\begin{equation}
    S_{raw}^\pm=\frac{dITF(\lambda)}{d\lambda}\bigg|_{\lambda=\lambda_{opt}^\pm}
\end{equation}

where,

\begin{equation}
    \lambda_{opt}^\pm=argmax\bigg\{\pm\frac{dITF(\lambda)}{d\lambda}\bigg\}
\end{equation}

which implies that in power limited measurements it might be beneficial to reduce the beam diameter in order to improve the measurement sensitivity. However, as discussed in ref. \cite{czuchnowski2021adaptive}, the raw optical sensitivity is not a good metric for comparing different FP sensor designs as it does not take into account the effects of laser related noise (e.g. shot noise) and shifts in the working point of the photodiode (PD) \cite{Czuchnowski:20}. Normalising the optical sensitivity corrects for the presence of these effects by incorporating the working point of the PD:

\begin{equation}
    S_{norm}^\pm=\frac{S_{raw}^\pm}{ITF(\lambda_{opt}^\pm)}
\end{equation}

When the normalisation is taken into account we can observe that larger beam diameters will lead to better performance of the sensor for both the falling and the rising edge (\textbf{Figure \ref{fig:1}E}). However, the performance of the sensor remains asymmetric with the rising edge being more sensitive.

\begin{figure*}[ht]
\includegraphics[width=16cm]{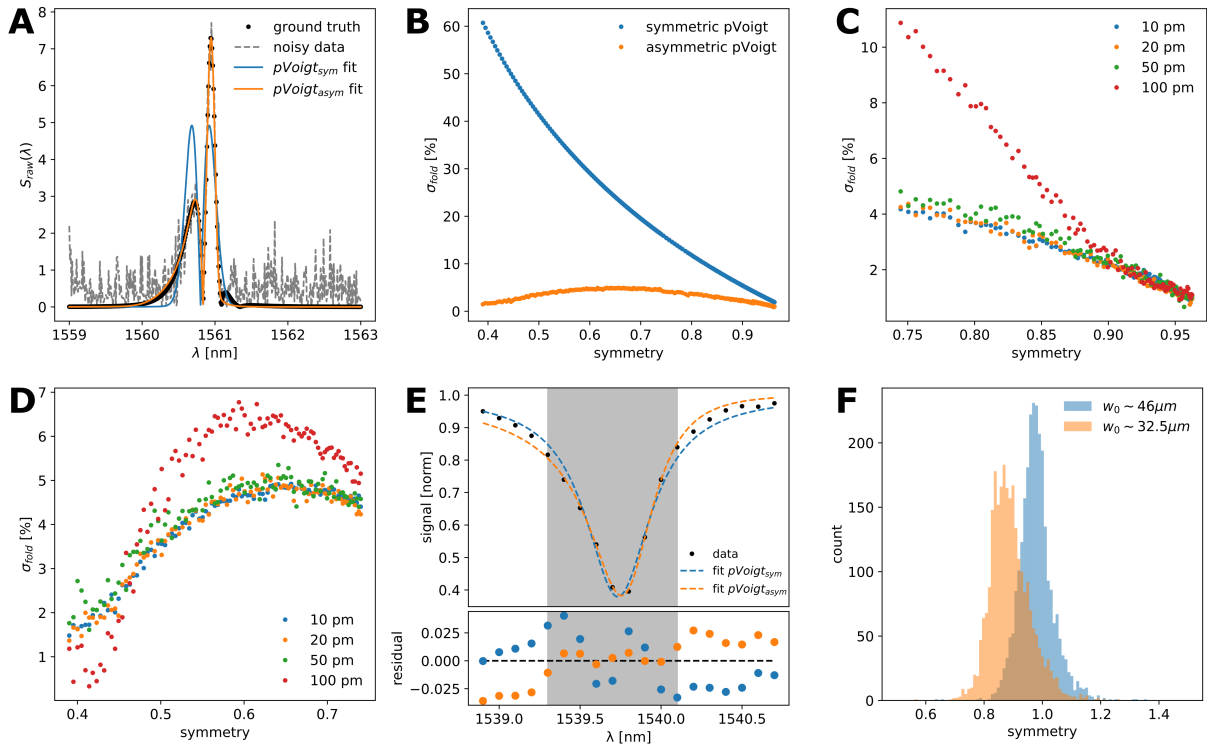}
\centering
\caption{\textbf{A} Exemplary fit to simulated data showing the superiority of the asymmetric pseudo-Voigt function ($pVoigt_{asym}$) in fitting highly asymmetric ITFs. The functions are fitted to simulated data based on \textbf{Equation 1} that includes shot noise to recreate experimental SNR conditions as well as provide the ground truth necessary to estimate the fitting error. \textbf{B} Dependence of the Log-fold error ($\sigma_{fold}$) on the ground truth asymmetry of the ITF. \textbf{C} $\sigma_{fold}$ dependence on the spectral sampling of the ITF showing high robustness for a large range of ITF sampling resolution. Simulation parameters: $w_0=40\ \mu m$, $R=0.98$, $l_0=20\ \mu m$. \textbf{D} $\sigma_{fold}$ dependence on the spectral sampling of the ITF showing high robustness for a large range of sampling resolution. Simulation parameters: $w_0=25\ \mu m$, $R=0.98$, $l_0=20\ \mu m$. \textbf{E} Exemplary fit of an experimentally acquired ITF comparing the symmetric and asymmetric pseudo-Voigt functions as well as their fitting residuals showing a clear improvement in the fit quality of the peak region (grey box). \textbf{F} ITF asymmetry quantification of a realistic FP pressure sensor ($R\sim0.98$, $l_0\sim20\ \mu m$, see \cite{czuchnowski2021adaptive} for details) using $pVoigt_{asym}$ fitting. For all panels: cavity thickness - $20 \mu m$, refractive index inside the cavity - 1.639, simulated wavelength range 1560-1562 nm.} 
\label{fig:2}
\end{figure*}

\subsection{Effects of mirror misalignment}

The simplest FP cavity imperfection to consider is mirror misalignment, which results in the cavity becoming wedged, because of the non-parallel orientation of the mirrors. As the FP pressure sensor is a solid cavity, these imperfections, which originate from the manufacturing process, are very difficult to correct. Thus it is important to evaluate the effects a cavity wedge angle has on the ITF asymmetry. 

To do so, we adopted a simple model based on a tilting frame of reference \cite{Varu:14}:

\begin{equation}
E_{wedge}(x,y,z,\lambda)=E(x',y,z',\lambda)
\end{equation}

where:

\begin{equation}
    z'=(x+x0)\sin(2k\alpha),
\end{equation}

\begin{equation}
    x'=(x+x0)\cos(2k\alpha)-x0,
\end{equation}

$x_0=l/tan(\alpha)$ and $\alpha$ is the wedge angle. We observed that large wedge angles limit the asymmetry (\textbf{Figure \ref{fig:1}F}). This is probably due to the fact that wedge angle, similarly to mirror reflectivity, is also capable of reducing the effective finesse of the FP pressure sensor (by limiting the spatial overlap of the reflections) which effectively reduces the asymmetry.

\subsection{Effects of optical aberrations}

It was previously described that optical aberrations lead to a generation of higher-order Laguerre-Gauss modes in Gaussian beams \cite{Mah:19}. These higher order modes are more likely to cause ITF asymmetry in the FP system because the Gouy phase is dependent on the beam order (\textbf{Equation \ref{eq:Gouy}}). To test this we used the theoretical framework developed in our previous work (see ref. \cite{Czuchnowski:20}) to evaluate the effects of beam aberrations on ITF asymmetry. We found that indeed stronger aberrations cause the ITF to become more asymmetric (\textbf{Figure \ref{fig:1}G}) which is in line with the intuitive prediction from analysing the Gouy phase term. This sheds additional light on the possible mechanism of the previously described ITF response to optical beam aberrations \cite{Czuchnowski:20}, which also behaves asymmetrically, with one edge being more resistant to the effects of Zernike-type aberrations.

%-----------------------------------
%	SECTION 1
%-----------------------------------

\section{Quantifying FPI sensitivity from experimental data}

Practical use of FP pressure sensors requires characterisation prior to measurement in order to determine the optimal bias wavelength. At the same time, comparison between different sensors requires quantification of the optical sensitivity. The recently described AO methods \cite{czuchnowski2021adaptive} for enhancing FP sensitivity require even more robust sensitivity quantification as it is used as a feedback readout for the AO optimisation. All of these applications require information to be extracted from an experimentally measured ITF. One of the most commonly used approaches is to fit the experimental ITF data with a generic line-shape function, such as a pseudo-Voigt function \cite{Buchmann:17,czuchnowski2021adaptive}:

\small
\begin{equation}
    V(\lambda;A,\gamma_0,\lambda_0)=fL(\lambda;A,\gamma_0,\lambda_0)+(1-f)G(\lambda;A,\gamma_0,\lambda_0)
\end{equation}
\normalsize

where $f$ is the relative ratio between the Lorentzian (\textit{L}) and Gaussian (\textit{G}) contributions to the line-shape \cite{stancik2008simple}:

\begin{equation}
    G(\lambda;A,\gamma_0,\lambda_0)=\frac{A}{\gamma_0}\sqrt{\frac{4\ln{2}}{\pi}}\exp{\bigg[-4\ln{2}\bigg(\frac{\lambda-\lambda_0}{\gamma_0}\bigg)^2\bigg]}
\end{equation}

and,

\begin{equation}
    L(\lambda;A,\gamma_0,\lambda_0)=\frac{2A}{\pi\gamma_0}\frac{1}{1+4[(\lambda-\lambda_0)/\gamma_0]^2}
\end{equation}

$A$ is the area under the curve, $\gamma_0$ is the FWHM of the line-shape and $\lambda_0$ is the peak position. This approach commonly used in spectroscopy works quite well for symmetric ITFs. However, it generally fails when fitting asymmetric ITFs since it assumes a symmetric line-shape. Here it can lead to an estimation error that is dependent on the degree of asymmetry (\textbf{Figure \ref{fig:2}A,B}):

\begin{equation}
    \sigma_{fold}\ [\%]=(\exp(|\ln(S_{fit}/S_{GT})|)-1)*100\%
\end{equation}

This can be corrected by allowing an asymmetrical line shape (see \textbf{Figure \ref{fig:2}A,B}), which can be realised in multiple ways. One of the simplest and most robust approaches is to allow the line-width to vary as a function of the wavelength using a sigmoid function \cite{stancik2008simple}:

\begin{equation}
    \gamma(\lambda;\gamma_0,\lambda_0,a)=\frac{2\gamma_0}{1+\exp[a(\lambda-\lambda_0)]}
\end{equation}

where the parameter $a$ determines the sigmoid slope and therefore controls the line asymmetry.

We have tested the robustness of this fitting by comparing the fit error with the sampling resolution of the spectral data used for fitting (\textbf{Figure \ref{fig:2}C,D}). We observed that the fitting precision does not improve beyond the sampling resolution of 50 pm for both of the tested conditions ($w_0=25/40\ \mu m$) but at the same time notice that the 40 \textmu m
beam (\textbf{Figure \ref{fig:2}C}) seems much more sensitive to undersampling. We conclude that this is predominantly due to the narrower ITF for a 40 \textmu m which will suffer more from undersampling than the broader ITF of a 25 \textmu m beam.

Finally, we have used the $pVoigt_{asym}$ model to fit experimentally measured ITFs (\textbf{Figure \ref{fig:2}E}, see ref. \cite{czuchnowski2021adaptive} for details) and experimentally quantified the ITF asymmetry dependence on the beam radius based on the fit parameters. The experimental results qualitatively follow the predictions from our simulations where smaller beam radius leads to stronger asymmetry  (\textbf{Figure \ref{fig:2}F}), corroborating the usefulness of our theoretical findings to experimental work.

\section{Discussion}

Fabry-Pérot pressure sensors are a rapidly developing type of optical ultrasound detectors for photoacoustic imaging. Accurately quantifying their sensitivity is important not only for optimising their design, but is also key for using advanced optical techniques (such as AO) to further improve their performance. Here, we discussed why ITF asymmetry arises in FP pressure sensors and how can it be affected by system imperfections (e.g. a wedged cavity or beam aberrations). Moreover we showed that implementing an asymmetric line-shape model allows for more accurate estimation of the optical sensitivity in theory as well as showed experimentally that the fitted line asymmetry follows qualitative predictions from our simulations.

We believe that taking into account the ITF asymmetry will allow to better quantify the performance of FP pressure sensors which will aid in pushing both AO based sensitivity enhancement further in future realizations as well as allow more standardised comparisons between FP sensors with different cavity parameters and application regimes.

\section*{Funding}
This work was supported by the European Molecular Biology Laboratory (EMBL) and the Chan Zuckerberg Initiative (Deep Tissue Imaging grant no. 2020-225346).

\section*{Disclosures}

The authors declare that there are no conflicts of interest related to this article.

\bibliographystyle{naturemag}  
\bibliography{biblio}

\newpage
\clearpage

\end{document}